\documentclass{PoS}

\title{Implications for compact stars of a soft nuclear equation of state from heavy-ion data}

\ShortTitle{Implications for compact stars of a soft nuclear equation of state from heavy-ion data}

\author{\speaker{Laura Tolos}\\
Instituto de Ciencias del Espacio (IEEC/CSIC) Campus Universitat Aut\`onoma de Barcelona, Facultat de Ci\`encies, Torre C5, E-08193 Bellaterra (Barcelona), Spain\\
Frankfurt Institute for Advanced Studies, Johann Wolfgang Goethe University, Ruth-Moufang-Str. 1
60438 Frankfurt am Main\\
E-mail:\email{tolos@ice.csic.es}}

\author{Irina Sagert \\
Department of Physics and Astronomy, Michigan State University, East Lansing, Michigan, 48824, USA}

\author{Debarati Chatterjee\\
Institut f\"ur Theoretische Physik, Ruprecht-Karls-Universit\"at, Philosophenweg 16, D-69120 Heidelberg, Germany}

\author{J\"urgen Schaffner-Bielich\\
Institut f\"ur Theoretische Physik, Ruprecht-Karls-Universit\"at, Philosophenweg 16, D-69120 Heidelberg, Germany}

\author{Christian Sturm\\
GSI Helmholtzzentrum f\"ur Schwerionenforschung GmbH, Planckstra\ss{}e~1, D-64291 Darmstadt, Germany\\
Institut f\"ur Kernphysik, J. W. Goethe Universit\"at, Max von Laue-Stra\ss{}e~1, D-60438 Frankfurt am Main, Germany}

\abstract{We study the implications on compact star properties of a soft nuclear equation of state determined from kaon production at subthreshold energies in heavy-ion collisions. On one hand, we apply these results to study 
radii and moments of inertia of light neutron stars. Heavy-ion data provides constraints on nuclear matter at densities relevant for those stars and, in particular, to the density dependence of the symmetry energy of nuclear matter. On the other hand,  we derive a limit for the highest allowed neutron star mass of three solar masses. For that purpose, we use the information on the nucleon potential obtained from the analysis of the heavy-ion data combined with causality on the nuclear equation of state.}

\FullConference{XII International Symposium on Nuclei in the Cosmos,\\
		August 5-12, 2012\\
		Cairns, Australia}
		
\begin{document}

\section{Introduction}
The study of neutron star properties such as their masses and radii has been a topic of high interest over the last decades.  Very recently, the measurement of the Shapiro delay for the millisecond pulsar PSR J1614-2230 has provided a new reliable limit for the highest known pulsar mass of $(1.97 \pm 0.04)\:$M$_\odot$ \cite{Demorest10}.  The value for the maximum mass is of major interest not only for astrophysics but also for nuclear physics as it is closely connected to the properties of dense nuclear matter, especially the stiffness of the nuclear matter equation of state (EoS).  Nuclear matter at finite baryon number densities is also subject of studies at heavy ion facilities such as Relativistic Heavy Ion Collider (RHIC) at Brookhaven, the future Facility for Antiproton and Ion Research (FAIR) at GSI, Darmstadt, and the Nuclotron-based Ion Collider facility (NICA) in Dubna, as well as the Facility for Rare Isotope Beams at the Michigan State University in East Lansing, USA. 

This work aims at studying the impact on the properties of compact stars of a soft nuclear matter EoS, as obtained from the KaoS experiment at GSI, Darmstadt, for baryon densities up to two-three times nuclear matter saturation density $n_0 \sim 0.17\:$fm$^{-3}$.  The K$^+$ meson production in nuclear collisions at subthreshold energies is used to  study the level of compression of dense matter, which, in turn, is controlled by the stiffness of nuclear matter through the nucleon potential U$_N$. The more attractive U$_N$ is, the higher is the produced  K$^+$  meson abundance, which therefore can serve as a probe for the stiffness of nuclear matter \cite{AichKo} . 

This paper is organized as follows. First, we review the results on K$^+$ meson production from the KaoS experiment and the need of a soft EoS to explain the observed data. Next, we study the relevance of the KaoS results for low mass neutron stars and their connection to the nuclear symmetry energy. This question arises  since light neutron stars can have interior densities in the same range as the ones probed by KaoS.  The last part of the paper is focused on the determination of the upper limit on neutron star maximum masses by using the information coming from the KaoS experiment and causality constraints. We conclude the paper with a discussion of the obtained limits for the highest possible neutron star mass with respect to neutron star observations, such as the PSR J1748-2021 pulsar \cite{Freire08}.

\section{A soft nuclear equation of state from KaoS experiment}

The measurements on K$^+$ meson production in nuclear collisions at subthreshold energies were performed with the Kaon Spectrometer (KaoS) at GSI. The beam energy dependence of the K$^+$ multiplicity ratio from Au+Au and C+C collisions at subthreshold energies of 0.8 to $1.5\:$GeV per nucleon was introduced as a sensitive and robust probe for the stiffness of nuclear matter \cite{Sturm01}.

In order to describe the experimental results, IQMD and RQMD (Isospin and Relativistic Quantum Molecular Dynamics) transport model calculations were done \cite{Fuchs01,Hartnack06}, applying a Skyrme type $U_N$ with two- and three-body forces. The detailed analysis showed that the K$^+$ multiplicity ratio from Au+Au and C+C collisions is rather insensitive to cross-sections, momentum dependence of in-medium potentials and transport models used while the experimental uncertainties cancel out in the double ratio \cite{Sturm01}. The transport calculations consistently demonstrate that the beam energy dependence of the kaon multiplicity ratio is described best by the nucleon potential with small repulsion. When applied to infinite isospin symmetric nuclear matter this corresponds to a soft nuclear EoS characterized by a stiffness parameter $K=9 \frac{dp}{dn} \left|_{n_0} \right. = 200\:$MeV \cite{Fuchs01}. 

The measured attractive nucleon potential and the corresponding soft nuclear EoS \cite{Sturm01,Fuchs01,Hartnack06} have to be tested on their compatibility with neutron star observations. In this work we want to analyze the implications of the results from the KaoS collaboration on compact stars.

\section{Radii and moments of inertia of low mass neutron stars: \\
density dependence of the nuclear symmetry energy }
 
The neutron star radii are a focus of major attention  \cite{Lattimer07,Ozel10,Steiner10,Suleimanov10,Horowitz03, Bauswein:2011tp,Bauswein:2012ya,Steiner:2012xt,Sala12} due to their implications for nuclear matter properties. Also moments of inertia seem to have captured the interest of the astrophysics community. Guillemot et al. have
presented a new technique to deduce limits on the moment of inertia of gamma-ray pulsars \cite{Guillemot12}. This analysis is based on the efficiency of the gamma ray emission from gamma-ray pulsars with respect to the total energy loss obtained from the pulsar spin-down.  Due to the fact that the energy emitted in gamma-rays cannot be larger than the total emitted energy, one can obtain a lower limit on the pulsar's moment of inertia. Also moments of inertia have been addressed in connection with  double pulsars \cite{lattimer2012}. Such studies could be substantially refined in the future with the Large Observatory For X-ray Timing (LOFT) \cite{Feroci:2012qh}.

Neutron star radii and moments of inertia are strongly connected to the symmetry energy \cite{Li08b}.  For our study, light neutron stars are of special interest, because their central densities can be in the same range as the ones probed by the KaoS experiment. Since the latter provides information on the stiffness of nuclear matter, the symmetry energy is the remaining uncertainty of the nuclear EoSs. 

For a consistency, we choose a Skyrme type EoS with a nucleon potential similar to the one which was used in the analysis of the KaoS data. The energy per baryon is written as \cite{Prakash88}:
\begin{eqnarray}
&&\frac{E}{A}= m_n\left(1-Y_p\right)+m_p Y_p+E_0 u^{\frac{2}{3}}+B \frac{u}{2}+D\frac{u^{\sigma}}{(\sigma+1)}
\nonumber\\
&+&\left(1-2Y_p\right)^2 \left[ \left(2^{\frac{2}{3}}-1\right)E_0\left(u^{\frac{2}{3}}-F(u) \right)+S_0 u^\gamma\right],
\label{pheneos}
\end{eqnarray}
whereas $u=n / n_0$ is the baryon number density and $E_0$ is the average kinetic energy of nuclear matter at $n_0$ and a proton fraction of $Y_p=0.5$. Two- and three-body forces, described by the terms $B$ and $D$, together with $\sigma$, are fitted to reproduce the binding energy per baryon $E(n_0,Y_p=0.5)=-16\:$MeV, the stiffness parameter $K$, and the saturation density $n_0$. While $S_0$ is varied between $28\:$MeV and $32\:$MeV, its density dependence is chosen as a power law with $u^{\gamma}$ where $\gamma=0.5-1.1$ \cite{Steiner10,Li06,Tsang09,Lattimer12}.

Eq.~(\ref{pheneos}) is used to describe matter in the neutron star core, while we incorporate an inner and outer crust  \cite{Negele73,Ruester06}. We first study non-rotating neutron stars with $M=1.25\:$M$_\odot$, in accord with the lightest pulsar masses  \cite{LatPrakBook}.
\begin{figure}
    \includegraphics[width=7.5cm]{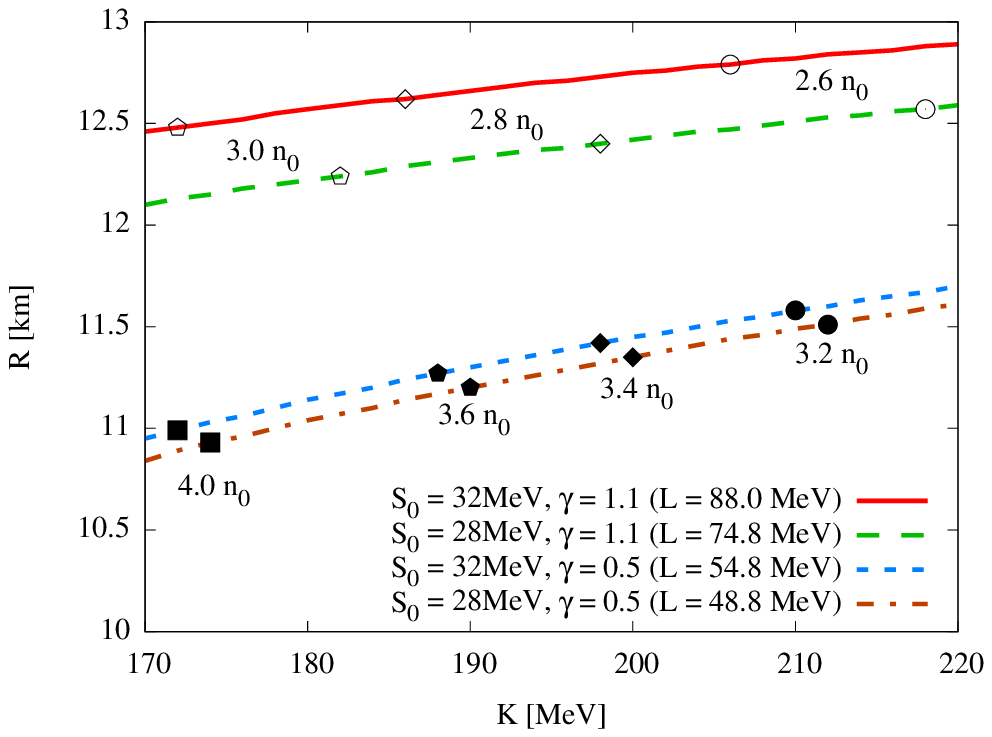}
    \hfill
    \includegraphics[width=7.5cm]{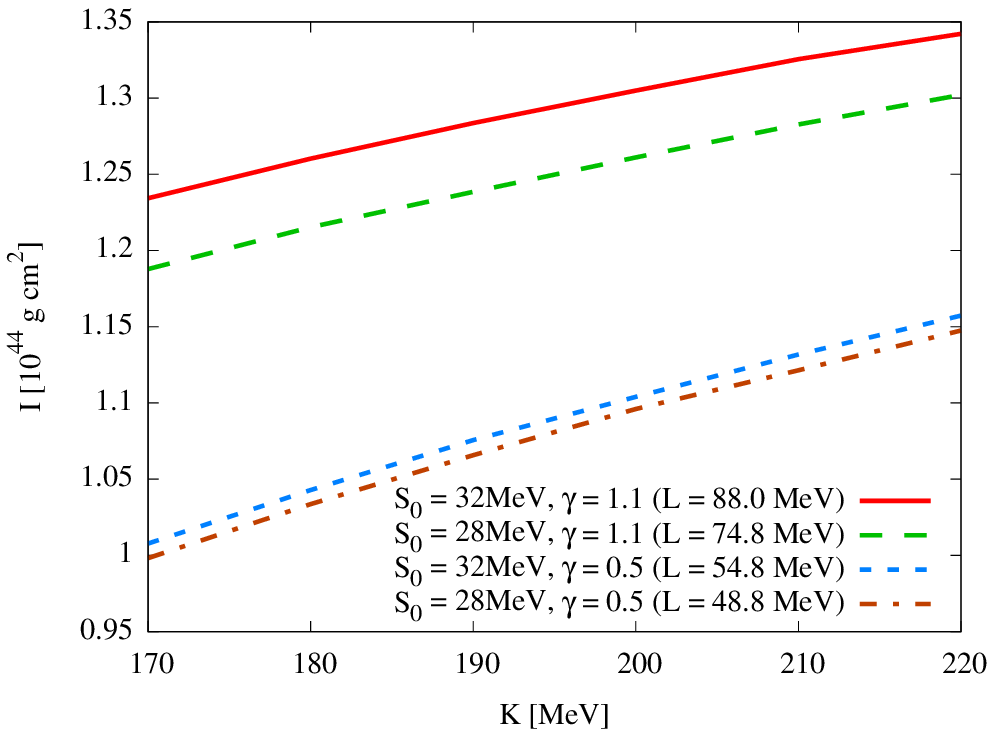}       
\caption{Radii $R$ and moments of inertia $I$ of non-rotating neutron star configurations with $1.25\:$M$_\odot$ as function of different slopes of the symmetry energy $L$. Central densities are given in units of the nuclear matter saturation density $n_0$. Taken from \cite{ours}.}
\label{cenden12}
\end{figure}
The radii and moments of inertia in dependence of $K$ are shown in Fig.~\ref{cenden12} for different symmetry energy configurations. It can be seen that both largely depend on the density dependence of the symmetry energy given by $\gamma$, i.e. $L=3 n_0 (d S(n_b) / d n_b)| n_0$. At $K \sim 200\:$MeV, stiff and soft symmetry energy configurations lead to a difference in the neutron star radius and moment of inertia of $\Delta R \sim (1-1.5)\:$km and $\Delta I \sim ( 2 \cdot 10^{43} - 2.5 \cdot 10^{43} ) \:$g cm$^2$, respectively. The central densities of the corresponding stars are in the range of $\lesssim 3.4\:n_0$, similar to the density region explored by the KaoS collaboration.

\begin{figure}
    \includegraphics[width=7.5cm]{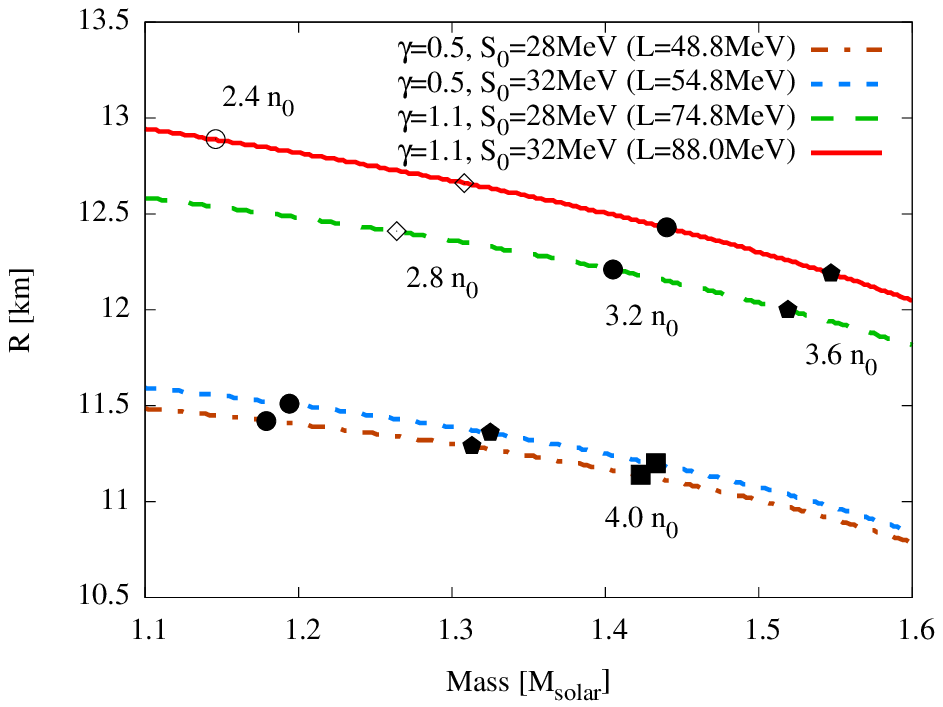}
    \hfill
    \includegraphics[width=7.5cm]{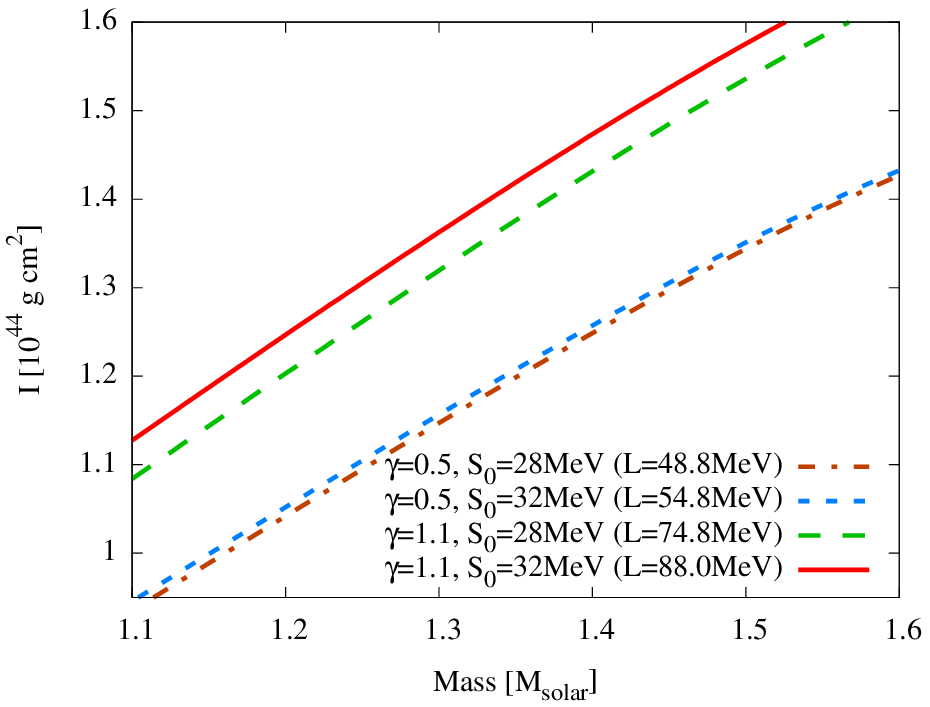}  
\caption{Radii and moments of inertia of light neutron stars for $K_0 \sim 200$MeV as function of their mass for different symmetry energy setups as in previous figure. Central densities are given in units of the nuclear matter saturation density $n_0$. Taken from \cite{ours}.}
\label{moi}
\end{figure}

In Fig.~\ref{moi}, we study the radii and moments of inertia of light neutron stars with masses in the range of $(1.1 - 1.6)\:$M$_\odot$ for $K_0 \sim 200\:$MeV and with equal configuration for the symmetry energy as in Fig.~\ref{cenden12}. It can be seen that especially neutron stars with $M \lesssim  1.3\:$M$_\odot$ have central densities which are low enough so  that the entire neutron star interior can be described by an EoS which is probed by the KaoS experiment. With the isospin symmetric part of the nuclear EoS determined by the latter, radius and moment of inertia measurements from light neutron stars could thereby distinguish between a soft and a stiff behavior of the symmetry energy. 

\section{Maximum mass for neutron stars}

The most massive stable neutron star configuration depends on the stiffness of nuclear matter. The softer the nuclear EoS, the lower is the maximum mass which can be reached before the star becomes unstable against collapse to a black hole. 

Ref.~\cite{Rhoades74} introduced the idea to use known properties of hadronic matter up to a critical density $n_{crit}$. At higher densities, the low density EoS is smoothly connected to the stiffest possible EoS allowed by causality. Previous studies showed that the highest possible mass can be expressed as \cite{LatPrakBook,Kalogera96,Wiringa88,Hartle78}
\begin{eqnarray}
M_{high} = 4.1 \: \mathrm{M}_\odot \left(\epsilon_{crit} / \epsilon_0  \right)^{-1/2} ,
\label{highmass}
\end{eqnarray}
with $\epsilon_{crit}$ being  the energy density taken at the transition density $n_{crit}$ from the known low density EoS to the stiffest one. In Refs.~\cite{Kalogera96,Hartle78,APR} an extrapolation of measurements on nuclei at $n \sim n_0$ to higher densities  were made and resulted in a maximum mass of  $M_{high} \sim 2.9\:$M$_\odot$ for $\epsilon_{crit} = 2\: \epsilon_0$.  KaoS data directly tests the EoS for densities $n \sim (2 - 3)\: n_0$ and justifies the choice of $n_{crit}$ in the same range. Consequently, we will calculate the highest possible masses with the restriction on $n_{crit}$ given by the KaoS data.

\begin{figure}
\centering
\includegraphics[width=7.5cm]{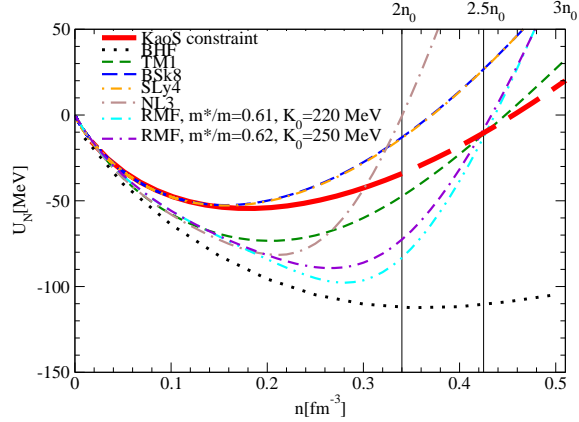}
\caption{The nucleon potential as a function of the baryon number density for different models. The vertical thin lines indicate different densities for the onset of the stiffest EoS. Taken from \cite{ours}.}
\label{fig:un}
\end{figure}

For $n_{crit}$ up to $(2-3) \: n_0$ the nucleon potential should fulfill the results from the analysis of the KaoS data.  For Skyrme type models this restriction corresponds to $K \leq 200\:$MeV. As seen in Fig.~\ref{fig:un}, we use two representative approaches for Skyrme models, BSk8 and SLy4 \cite{Goriely05,Chabanat98}. For Relativistic Mean Field (RMF) EoSs, the stiffness of high density nuclear matter is determined by the nucleon effective mass $m^*$ at $n_0$  \cite{BogutaStoecker}. Consequently, the corresponding $U_N$ is chosen by varying $m^*$ for a given compression modulus $K_0$, so as to obtain a nucleon potential which is similar to or even more attractive than the Skyrme parametrization within the density limits. A more attractive $U_N$ at supra-saturation densities allows a higher compression of matter for the same bombarding energy, which enhances multiple-step processes in subthreshold kaon production and is in line with KaoS results.  Fig.~\ref{fig:un} also shows that the TM1 \cite{Sugahara94} parametrization as well as the Brueckner-Hartree-Fock approximation (BHF) \cite{isaac,schulze,zuo,bombaci} fulfill this requirement. Schemes, such as BSk8 \cite{Goriely05}, SLy4 \cite{Chabanat98} or NL3 \cite{lalazissis}, produce nucleon potentials which are more repulsive than the Skyrme benchmark in the density region of interest.

\begin{figure}
  \centering
    \includegraphics[width=7.5cm]{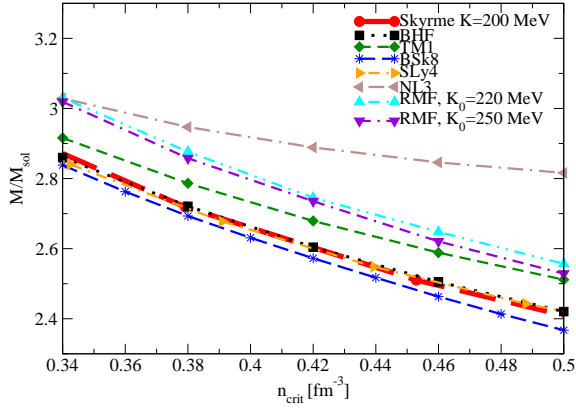}
\caption{Highest possible masses for compact stars for different EoS as a function of the critical density  for the onset of the stiffest possible EoS.
Taken from \cite{ours}.}
\label{highestm}
\end{figure}

Fig.~\ref{highestm} shows the highest possible neutron star masses at $n_{crit} \sim (2-3) \: n_0$ for the discussed EoSs. Note that the maximum mass does not depend on the asymmetry at low densities but on the critical density. The higher $n_{crit}$ is, the later is the onset of the stiffest EoS in the star's interior. Consequently, less mass is supported and the value for the highest mass decreases. For $n_{crit} \sim 2 \: n_0$, the star is dominated by the stiffest EoS and therefore reaches masses of up to $3\:$M$_\odot$. Smaller maximum masses  of $M_{high} \sim 2.4\:$M$_\odot$ are obtained for the upper limit of $n_{crit} \simeq 3 \: n_0$.  However, this lower bound depends on the EoS studied. Moreover, it can be observed that a lower bound of $2.8\:$M$_\odot$ as given by the NL3 EoS is not compatible with constraints from KaoS data. From Fig.~\ref{highestm} we can conclude that a pulsar mass of $2.7\:$M$_\odot$ as found for PSR J1748-2021 by \cite{Freire08} is marginally compatible with a soft EoS and requires a prompt transition from a soft EoS to the stiffest possible for a density around $(2.2 - 2.5)\:n_0$.

\section{Summary}

We have studied the properties of low mass neutron stars, such as radius and moments of inertia, using  K$^+$ multiplicities from heavy-ion collisions at GSI. Light mass neutron stars can have interior densities in the same range as the ones probed by KaoS experiment. Thus, these experimental results serve to constrain the stiffness of the nuclear EoS for  low mass stars. The radii and moments of inertia of low mass neutron stars are then sensitive to the density dependence of the symmetry energy, which is the remaining uncertainty of the nuclear EoSs. We find that light neutron stars with $M \lesssim 1.25\:$M$_\odot$ are well suited for future radius measurements, such as the LOFT mission.

Moreover, we apply the KaoS results up to densities of $2\: n_0 \leq n_{crit} \leq 3 \: n_0$ and introduce the stiffest possible EoS for $n > n_{crit}$ to calculate the highest allowed maximum mass.  In this way,  we test whether a soft nuclear EoS up to $\sim 3 \: n_0$ is compatible with tentatively massive neutron stars such as PSR J1748-2021. The KaoS results confirm the previous theoretical estimation for the highest possible neutron star mass of $\sim 3\:$M$_{\odot}$, estimation based on an extrapolation to supra-saturation densities of results for nuclei at $n \sim n_0$. A pulsar mass of $2.7\:$M$_{\odot}$ is not excluded by the KaoS data, but requires the onset of the stiffest possible EoS already at $n_{crit} \sim 2.2-2.5 \: n_0$.  Future experiments, such as FAIR at GSI, aim at probing matter at densities beyond  $3\:n_0$ and, therefore,  giving further constraints on maximum masses.

\acknowledgments{
 D.C. and I.S. are thankful to the Alexander von Humboldt foundation. I.S. acknowledges the support of the High Performance Computer Center and the Institute for Cyber-Enabled Research at Michigan State University. L.T. is supported from MICINN under FPA2010-16963 and RyC2009, and from  FP7 under PCIG09-GA-2011-291679. J.S.-B. is supported by the DFG through the Heidelberg Graduate School for Fundamental Physics and by the BMBF grant 06HD9127. }


\begin{thebibliography}{99}

%\cite{Demorest:2010bx}
%\bibitem{Demorest:2010bx} 
 \bibitem{Demorest10}
  P.~Demorest, T.~Pennucci, S.~Ransom, M.~Roberts and J.~Hessels,
  %``Shapiro Delay Measurement of A Two Solar Mass Neutron Star,''
  Nature {\bf 467}, 1081 (2010)
%  [arXiv:1010.5788 [astro-ph.HE]].
  %%CITATION = ARXIV:1010.5788;%%

%\cite{Aichelin:1986ss}
%\bibitem{Aichelin:1986ss} 
 \bibitem{AichKo}
  J.~Aichelin and C.~M.~Ko,
  %``Subthreshold Kaon Production as a Probe of the Nuclear Equation of State,''
  Phys.\ Rev.\ Lett.\  {\bf 55}, 2661 (1985).
  %%CITATION = PRLTA,55,2661;%%

%\cite{Freire:2007jd}
%\bibitem{Freire:2007jd} 
  \bibitem{Freire08}
  P.~C.~C.~Freire, S.~M.~Ransom, S.~Begin, I.~H.~Stairs, J.~W.~T.~Hessels, L.~H.~Frey and F.~Camilo,
  %``Eight New Millisecond Pulsars in NGC 6440 and NGC 6441,''
    %Submitted to: Astrophys.J.
  Astrophys. J. {\bf 675}, 670 (2008)
  %[arXiv:0711.0925 [astro-ph]].
  %%CITATION = ARXIV:0711.0925;%%

%\cite{Sturm:2000dm}
%\bibitem{Sturm:2000dm} 
 \bibitem{Sturm01}
  C.~T.~Sturm {\it et al.}  [KAOS Collaboration],
  %``Evidence for a soft nuclear equation of state from kaon production in heavy ion collisions,''
  Phys.\ Rev.\ Lett.\  {\bf 86}, 39 (2001)
  %[nucl-ex/0011001].
  %%CITATION = NUCL-EX/0011001;%%

%\cite{Fuchs:2000kp}
%\bibitem{Fuchs:2000kp} 
 \bibitem{Fuchs01}
  C.~Fuchs, A.~Faessler, E.~Zabrodin and Y.~-M.~Zheng,
  %``Probing the nuclear equation of state by K+ production in heavy ion collisions,''
  Phys.\ Rev.\ Lett.\  {\bf 86}, 1974 (2001)
 % [nucl-th/0011102].
  %%CITATION = NUCL-TH/0011102;%%

%\cite{Hartnack:2005tr}
%\bibitem{Hartnack:2005tr} 
 \bibitem{Hartnack06}
  C.~.Hartnack, H.~Oeschler and J.~Aichelin,
  %``Hadronic matter is soft,''
  Phys.\ Rev.\ Lett.\  {\bf 96}, 012302 (2006)
  %[nucl-th/0506087].
  %%CITATION = NUCL-TH/0506087;%%
  
  %\cite{Lattimer:2006xb}
%\bibitem{Lattimer:2006xb} 
 \bibitem{Lattimer07}
  J.~M.~Lattimer and M.~Prakash,
  %``Neutron Star Observations: Prognosis for Equation of State Constraints,''
  Phys.\ Rept.\  {\bf 442}, 109 (2007)
 % [astro-ph/0612440].
  %%CITATION = ASTRO-PH/0612440;%%

%\cite{Ozel:2010fw}
%\bibitem{Ozel:2010fw} 
  \bibitem{Ozel10}
  F.~Ozel, G.~Baym and T.~Guver,
  %``Astrophysical Measurement of the Equation of State of Neutron Star Matter,''
  Phys.\ Rev.\ D {\bf 82}, 101301 (2010)
 % [arXiv:1002.3153 [astro-ph.HE]].
  %%CITATION = ARXIV:1002.3153;%%

%\cite{Steiner:2010fz}
%\bibitem{Steiner:2010fz} 
  \bibitem{Steiner10}
  A.~W.~Steiner, J.~M.~Lattimer and E.~F.~Brown,
  %``The Equation of State from Observed Masses and Radii of Neutron Stars,''
  Astrophys.\ J.\  {\bf 722}, 33 (2010)
  %[arXiv:1005.0811 [astro-ph.HE]].
  %%CITATION = ARXIV:1005.0811;%%

%\cite{Suleimanov:2010ij}
%\bibitem{Suleimanov:2010ij} 
 \bibitem{Suleimanov10}
  V.~Suleimanov, J.~Poutanen, M.~Revnivtsev and K.~Werner,
  %``Model atmospheres of X-ray bursting neutron stars,''
  AIP Conf.\ Proc.\  {\bf 1379}, 197 (2011)
  %[arXiv:1010.0151 [astro-ph.HE]].
  %%CITATION = ARXIV:1010.0151;%%

%\cite{Carriere:2002bx}
%\bibitem{Carriere:2002bx} 
  \bibitem{Horowitz03}
  J.~Carriere, C.~J.~Horowitz and J.~Piekarewicz,
  %``Low mass neutron stars and the equation of state of dense matter,''
  Astrophys.\ J.\  {\bf 593}, 463 (2003)
  %[nucl-th/0211015].
  %%CITATION = NUCL-TH/0211015;%%
  
  %\cite{Bauswein:2011tp}
\bibitem{Bauswein:2011tp} 
  A.~Bauswein and H.~-T.~.Janka,
  %``Measuring neutron-star properties via gravitational waves from binary mergers,''
  Phys.\ Rev.\ Lett.\  {\bf 108}, 011101 (2012)
 % [arXiv:1106.1616 [astro-ph.SR]].
  %%CITATION = ARXIV:1106.1616;%%

%\cite{Bauswein:2012ya}
\bibitem{Bauswein:2012ya} 
  A.~Bauswein, H.~-T.~Janka, K.~Hebeler and A.~Schwenk,
  %``Equation-of-state dependence of the gravitational-wave signal from the ring-down phase of neutron-star mergers,''
  arXiv:1204.1888 [astro-ph.SR].
  %%CITATION = ARXIV:1204.1888;%%

%\cite{Steiner:2012xt}
\bibitem{Steiner:2012xt} 
  A.~W.~Steiner, J.~M.~Lattimer and E.~F.~Brown,
  %``The Neutron Star Mass-Radius Relation and the Equation of State of Dense Matter,''
  arXiv:1205.6871 [nucl-th].
  %%CITATION = ARXIV:1205.6871;%%


%\cite{Sala:2012ng}
%\bibitem{Sala:2012ng} 
 \bibitem{Sala12}
  G.~Sala, F.~Haberl, J.~Jose, A.~Parikh, R.~Longland, L.~C.~Pardo and M.~Andersen,
  %``Constraints on the mass and radius of the accreting neutron star in the Rapid Burster,''
  Astrophys.\ J.\  {\bf 752}, 158 (2012)
 % [arXiv:1204.3627 [astro-ph.HE]].
  %%CITATION = ARXIV:1204.3627;%%
  
  
  %\cite{Guillemot:2012ne}
%\bibitem{Guillemot:2012ne} 
  \bibitem{Guillemot12}
  L.~Guillemot, P.~C.~C.~Freire, I.~Cognard, T.~J.~Johnson, Y.~Takahashi, J.~Kataoka, G.~Desvignes and F.~Camilo {\it et al.},
  %``Discovery of the millisecond pulsar PSR J2043+1711 in a Fermi source with the Nancay Radio Telescope,''
  Mon.\ Not.\ Roy.\ Astron.\ Soc.\  {\bf 422}, 1294 (2012)
%  [arXiv:1202.1128 [astro-ph.HE]].
  %%CITATION = ARXIV:1202.1128;%%


\bibitem{lattimer2012}
%Constraining the equation of state with moment of inertia measurements. 
J. M. Lattimer and B. F. Schutz, 
Astrophys.~J.~{\bf 629}, 979 (2005)


%\cite{Feroci:2012qh}
\bibitem{Feroci:2012qh} 
  M.~Feroci, J.~W.~d.~Herder, E.~Bozzo, D.~Barret, S.~Brandt, M.~Hernanz, M.~van der Klis and M.~Pohl {\it et al.},
  %``LOFT: the Large Observatory For X-ray Timing,''
  arXiv:1209.1497 [astro-ph.IM].
  %%CITATION = ARXIV:1209.1497;%%
  
  %\cite{Li:2007pw}
%\bibitem{Li:2007pw} 
  \bibitem{Li08b}
  B.~-A.~Li, L.~-W.~Chen, C.~M.~Ko, P.~G.~Krastev, A.~W.~Steiner and G.~-C.~Yong,
  %``Constraining properties of neutron stars with heavy-ion reactions in terrestrial laboratories,''
  J.\ Phys.\ G G {\bf 35}, 014044 (2008)
  %[arXiv:0705.2999 [nucl-th]].
  %%CITATION = ARXIV:0705.2999;%%

%\cite{Prakash:1988md}
%\bibitem{Prakash:1988md} 
 \bibitem{Prakash88}
  M.~Prakash, T.~L.~Ainsworth and J.~M.~Lattimer,
  %``Equation of state and the maximum mass of neutron stars,''
  Phys.\ Rev.\ Lett.\  {\bf 61}, 2518 (1988).
  %%CITATION = PRLTA,61,2518;%%

%\cite{Li:2005sr}
%\bibitem{Li:2005sr} 
  \bibitem{Li06}
  B.~-A.~Li and A.~W.~Steiner,
  %``Constraining the radii of neutron stars with terrestrial nuclear laboratory data,''
  Phys.\ Lett.\ B {\bf 642}, 436 (2006)
 % [nucl-th/0511064].
  %%CITATION = NUCL-TH/0511064;%%

%\cite{Tsang:2008fd}
%\bibitem{Tsang:2008fd} 
 \bibitem{Tsang09}
  M.~B.~Tsang, Y.~Zhang, P.~Danielewicz, M.~Famiano, Z.~Li, W.~G.~Lynch and A.~W.~Steiner,
  %``Constraints on the density dependence of the symmetry energy,''
  Phys.\ Rev.\ Lett.\  {\bf 102}, 122701 (2009)
  [Int.\ J.\ Mod.\ Phys.\ E {\bf 19}, 1631 (2010)]
 % [arXiv:0811.3107 [nucl-ex]].
  %%CITATION = ARXIV:0811.3107;%%


%\cite{Lattimer:2012xj}
%\bibitem{Lattimer:2012xj} 
  \bibitem{Lattimer12}
  J.~M.~Lattimer and Y.~Lim,
  %``Constraining the Symmetry Parameters of the Nuclear Interaction,''
  arXiv:1203.4286 [nucl-th].
  %%CITATION = ARXIV:1203.4286;%%
  
  %\cite{Negele:1971vb}
%\bibitem{Negele:1971vb} 
 \bibitem{Negele73}
  J.~W.~Negele and D.~Vautherin,
  %``Neutron star matter at subnuclear densities,''
  Nucl.\ Phys.\ A {\bf 207}, 298 (1973).
  %%CITATION = NUPHA,A207,298;%%

  %\cite{Ruester:2005fm}
%\bibitem{Ruester:2005fm} 
  \bibitem{Ruester06}
  S.~B.~Ruester, M.~Hempel and J.~Schaffner-Bielich,
  %``The outer crust of non-accreting cold neutron stars,''
  Phys.\ Rev.\ C {\bf 73}, 035804 (2006)
 % [astro-ph/0509325].
  %%CITATION = ASTRO-PH/0509325;%%
  
  %\cite{Rhoades:1974fn}
%\bibitem{Rhoades:1974fn} 
\bibitem{Rhoades74}  
  C.~E.~Rhoades, Jr. and R.~Ruffini,
  %``Maximum mass of a neutron star,''
  Phys.\ Rev.\ Lett.\  {\bf 32}, 324 (1974).
  %%CITATION = PRLTA,32,324;%%
  
  \bibitem{LatPrakBook}
J. M. Lattimer and M. Prakash, From Nuclei to Stars: Festschrift in Honor of Gerald Brown,
(World Scientific Publishing Co. Pt. Ltd., 275, 2001)

\bibitem{ours}
%\cite{Sagert:2011kf}
%\bibitem{Sagert:2011kf} 
  I.~Sagert, L.~Tolos, D.~Chatterjee, J.~Schaffner-Bielich and C.~Sturm,
  %``Soft nuclear equation-of-state from heavy-ion data and implications for compact stars,''
 Phys. Rev. C {\bf 86}, 045802 (2012)
  %%CITATION = ARXIV:1111.6058;%%

%\cite{Kalogera:1996ci}
%\bibitem{Kalogera:1996ci} 
  \bibitem{Kalogera96}
  V.~Kalogera and G.~Baym,
  %``The maximum mass of a neutron star,''
  Astrophys.\ J.\  {\bf 470}, L61 (1996)
  %[astro-ph/9608059].
  %%CITATION = ASTRO-PH/9608059;%%

%\cite{Wiringa:1988tp}
%\bibitem{Wiringa:1988tp} 
 \bibitem{Wiringa88}
  R.~B.~Wiringa, V.~Fiks and A.~Fabrocini,
  %``Equation of state for dense nucleon matter,''
  Phys.\ Rev.\ C {\bf 38}, 1010 (1988).
  %%CITATION = PHRVA,C38,1010;%%

\bibitem{Hartle78}
	J. B. {Hartle}, Phys. Rep. {\bf 46}, 201 (1978)



%\cite{Akmal:1998cf}
%\bibitem{Akmal:1998cf} 
  \bibitem{APR}
  A.~Akmal, V.~R.~Pandharipande and D.~G.~Ravenhall,
  %``The Equation of state of nucleon matter and neutron star structure,''
  Phys.\ Rev.\ C {\bf 58}, 1804 (1998)
 % [nucl-th/9804027].
  %%CITATION = NUCL-TH/9804027;%%
  
 \bibitem{Goriely05}
S. {Goriely} and M. {Samyn} and J.~M. {Pearson} and M. {Onsi},
Nuc. Phys. A {\bf 750}, 425 (2005)


%\cite{Chabanat:1997un}
%\bibitem{Chabanat:1997un} 
  \bibitem{Chabanat98} 
  E.~Chabanat, P.~Bonche, P.~Haensel, J.~Meyer and R.~Schaeffer,
  %``A Skyrme parametrization from subnuclear to neutron star densities. 2. Nuclei far from stablities,''
  Nucl.\ Phys.\ A {\bf 635}, 231 (1998).
  %%CITATION = NUPHA,A635,231;%%

%\cite{Boguta:1981px}
%\bibitem{Boguta:1981px} 
 \bibitem{BogutaStoecker}
  J.~Boguta and H.~Stoecker,
  %``Systematics Of Nuclear Matter Properties In A Nonlinear Relativistic Field Theory,''
  Phys.\ Lett.\ B {\bf 120}, 289 (1983).
  %%CITATION = PHLTA,B120,289;%%

%\cite{Sugahara:1993wz}
%\bibitem{Sugahara:1993wz} 
 \bibitem{Sugahara94}
  Y.~Sugahara and H.~Toki,
  %``Relativistic mean field theory for unstable nuclei with nonlinear sigma and omega terms,''
  Nucl.\ Phys.\ A {\bf 579}, 557 (1994).
  %%CITATION = NUPHA,A579,557;%%

%\cite{Schulze:2006vw}
%\bibitem{Schulze:2006vw} 
 \bibitem{isaac}
  H.~-J.~Schulze, A.~Polls, A.~Ramos and I.~Vidana,
  %``Maximum mass of neutron stars,''
  Phys.\ Rev.\ C {\bf 73}, 058801 (2006).
  %%CITATION = PHRVA,C73,058801;%%

%\cite{Schulze:2011zza}
%\bibitem{Schulze:2011zza} 
 \bibitem{schulze}
  H.~-J.~Schulze and T.~Rijken,
  %``Maximum mass of hyperon stars with the Nijmegen ES C-08 model,''
  Phys.\ Rev.\ C {\bf 84}, 035801 (2011).
  %%CITATION = PHRVA,C84,035801;%%
  
  %\cite{Zuo:2001bd}
%\bibitem{Zuo:2001bd} 
 \bibitem{zuo}
  W.~Zuo, I.~Bombaci and U.~Lombardo,
  %``Asymmetric nuclear matter from extended Brueckner-Hartree-Fock approach,''
  Phys.\ Rev.\ C {\bf 60}, 024605 (1999)
%  [nucl-th/0102035].
  %%CITATION = NUCL-TH/0102035;%%

%\cite{Bombaci:1991zz}
%\bibitem{Bombaci:1991zz} 
  \bibitem{bombaci}
  I.~Bombaci and U.~Lombardo,
  %``Asymmetric nuclear matter equation of state,''
  Phys.\ Rev.\ C {\bf 44}, 1892 (1991).
  %%CITATION = PHRVA,C44,1892;%%

%\cite{Lalazissis:1996rd}
%\bibitem{Lalazissis:1996rd} 
  \bibitem{lalazissis}
  G.~A.~Lalazissis, J.~Konig and P.~Ring,
  %``A New parametrization for the Lagrangian density of relativistic mean field theory,''
  Phys.\ Rev.\ C {\bf 55}, 540 (1997)
 % [nucl-th/9607039].
  %%CITATION = NUCL-TH/9607039;%%



\end{thebibliography}
\end{document}